# The key role of smooth impurity potential in formation of hole spectrum for p-Ge/Ge$_{1-x}$Si$_x$ heterostructures in the quantum Hall regime

Yu G Arapov*, G A Alshanskii, G I Harus, V N Neverov, N G Shelushinina, M V Yakunin, O A Kuznetsov[#]

Institute of Metal Physics RAS, Ekaterinburg, GSP-170, 620219, Russia
[#]Scientific-Research Institute at Nizhnii Novgorod State University, Russia

**Abstract.** We have measured the temperature ($0.1 \leq T \leq 15$ K) and magnetic field ($0 \leq B \leq 12$ T) dependences of longitudinal and Hall resistivities for the p-Ge$_{0.93}$Si$_{0.07}$/Ge multilayers with different Ge layer widths $10 \leq d_w \leq 38$ nm and hole densities $p_s = (1 \div 5) \cdot 10^{11}$ cm$^{-2}$. Two models for the long-range random impurity potential (the model with randomly distributed charged centers located outside the conducting layer and the model of the system with a spacer) are used for evaluation of the impurity potential fluctuation characteristics: the random potential amplitude, nonlinear screening length in vicinity of integer filling factors $n = 1$ and $n = 2$ and the background density of state (DOS). The described models are suitable for explanation of the unusually high value of DOS at $n = 1$ and $n = 2$, in contrast to the short-range impurity potential models. For half-integer filling factors the linear temperature dependence of the effective QHE plateau-to-plateau transition width $n_0(T)$ is observed in contrast to scaling behavior for systems with short-range disorder. The finite $T \to 0$ width of QHE transitions may be due to an effective low temperature screening of smooth random potential owing to Coulomb repulsion of electrons.

## 1. Introduction

For the two-dimensional (2D) systems with the electron gas of high degeneracy the magnetic field dependence of the Hall component of magnetoresistance tensor $r_{xy}(B)$ is a set of plateaux with universal values of $r_{xy}^{(i)} = h/(ie^2)$ where $i$ is an integer [1]. The adjacent quantum Hall plateaus of a width $\Delta B_i$ are divided by a narrow intervals of magnetic field $dB_i$ where $r_{xy}$ is jumping from one plateau to another with $dr_{xy}/dB \gg dr_{xy}^{classical}/dB = (nec)^{-1}$. When temperature is lowered the width of the intervals $\Delta B_i$ increases while that of $dB_i$ decreases so that the derivative $dr_{xy}/dB \cong h/[i(i+1)e^2 dB_i]$ becomes larger.

The nature of the quantum Hall effect (QHE) has occurred to be closely linked with a phenomenon of electron localization in 2D-disorder system at quantizing magnetic fields [2, 3]. Laughlin [2] and Halperin [3] showed that, for the QHE to exist, narrow bands of extended states must be present close to the center of each of the Landau subbands provided that all the other states are localized. When the magnetic field values are in the plateau regions ($\Delta B_i$ intervals) the system is in the localized regime and temperature dependence of dissipative conductivity $s_{xx}$ (and resistivity $r_{xx} \cong s_{xx}/s_{xy}^2$) is of exponential character, $s_{xx}(T) \to 0$ as $T \to 0$ [4,5]. If the magnetic field is in the plateau–plateau (PP) transition region ($dB_i$ intervals) the Fermi level passes through the narrow strip of extended states at a Landau level (LL) center. The system behaves itself as a metal with non-zero conductivity at $T \to 0$ and the peak-like form of $s_{xx}(B)$ dependence.

An analysis of temperature dependence of $r_{xx}$ and $r_{xy}$ both in the plateau and in the PP transition regions allows extracting such parameters of electron spectrum as the energy separation between the adjacent LL's, the relative fractions of localized and extended states, the density of localized states and the width of extended state bands. The experimental reconstruction of energy spectrum is especially actual for p-type systems with complex valence band spectrum. In such a system the LL picture is not determined only by the cyclotron energy $\hbar\omega_c$ with a given effective mass as that for n-type system with simple parabolic conduction band.

* To whom correspondence should be addressed.
e-mail: arapov@imp.uran.ru

In this paper we report on magnetoresistance investigations just for p-type system, namely, for multilayer Ge/Ge$_{1-x}$Si$_x$ heterostructures with hole conduction over the germanium layers. A brief description of samples and experimental details will be done in section 2. The results for QHE plateau regions will be presented in section 3.1 where the width of inter-LL mobility gaps and the background density of localized states will be evaluated from the analysis of activated magnetoresistivity. Two models of random impurity potential will be used for evaluation of the impurity potential fluctuation parameters: random potential amplitude and the nonlinear screening length in a vicinity of integer filling factors (FF's). In section 3.2 the data for QHE PP transition regions will be reported and the temperature dependences both of the width of extended state band and of the conductivity on these states will be extracted and analyzed in terms of critical phenomena theory. The influence of Coulomb interaction on smooth disorder potential screening will be discussed. The concluding remarks will be given in section 4.

## 2. The Ge/Ge$_{1-x}$Si$_x$ heterostructures and experimental procedure

The multilayer, selectively doped p-Ge/Ge$_{1-x}$Si$_x$ ($x = 0.07$) heterostructures studied here contain from 15 to 30 periods, with Ge and Ge$_{1-x}$Si$_x$ layers of width $d_w = (200 \div 230)$Å. The Ge layers are undoped, but the GeSi layers are doped with boron in such a way that spacers $d_s$ about 50Å thick remain between the doped part of the solid solution (of $d_a \cong 100$Å width) and the germanium layers. The top of valence band in the Ge layer is located higher in energy than that in the Ge$_{1-x}$Si$_x$ layer. As a result, holes from the doped part of the solid solution pass into the Ge layers. The growth methods and other properties of the p-Ge/Ge$_{1-x}$Si$_x$ heterostructures are described in more detail in earlier papers (see [6] and references therein).

Samples in the shape of Hall bars with a size of 0.27cm × 0.05cm were fabricated for the measurements. The measurements were carried out in an Oxford superconducting solenoid in magnetic fields up to 12T in a temperature interval $T = (0.1 \div 15)$K. The hole concentration was determined from Hall measurements in a weak magnetic field and from the period of the Shubnikov-de Haas oscillations for large LL numbers. Results for the samples 1124b3, 1125a7 and 1123a6 with hole concentration $p = (2.4 \div 2.6) \cdot 10^{11}$cm$^{-2}$ and mobility $m_p = (1.1 \div 1.7) \cdot 10^4$ cm$^2$/Vs are reported here.

We made a certain conclusion that these are the boron ionized impurities (with concentration $N \sim 10^{17}$cm$^{-3}$) in the central parts of GeSi layers that limit the hole mobility in the Ge layers. For that we calculated mobility of carriers in a quantum well where only the lowest quantized energy level is populated [7]. The scattering arises from remote charged impurities via the long-range Coulomb interaction. The simplest results may be obtained in a delta-function approximation for both the carrier and the impurity probability density. The plane with 2D hole gas is assumed to be separated from the impurity plane by an effective spacer $s = d_s + d_w/2 + d_a/2 = 200$Å. A finite mobility for remote ionized impurity scattering of the order of $3 \cdot 10^3$ cm$^2$/Vs is found even in the absence of screening [7] unlike the case for bulk semiconductors. In order to take into account the effect of screening we use the strictly 2D dielectric function $æ(q) = æ(1+q/q_s)$ where $q_s$ is the inverse screening radius. Then the calculated value of $m_p \cong 1.5 \cdot 10^4$ cm$^2$/Vs is obtained in accordance with experimental values of hole mobilities.

## 3. Experimental results and discussion

### 3.1. QHE plateau regions

#### 3.1.1. The density of states in the mobility gap

The appearance of quantum plateaux in the $r_{xy}(B)$ dependences with vanishing values of $r_{xx}$ is now commonly accepted to be caused by the existence of disorder-induced mobility gaps in the density of states (DOS) of a 2D-system. When the Fermi level is settled down in the gap, the thermally activated behavior of $r_{xx}$ (or $s_{xx}$) is observed due to excitation of electrons into very narrow bands of extended states centered at LL energies $E_N$. The DOS in mobility gaps may be evaluated from the data on activation energy $E_A$ as a function of the LL filling factor $n = n/n_B$ ($n$ is the electron density, $n_B = eB/hc$). The filling factor can be tuned by the change of either a carrier density [8] or a magnetic field [4, 5, 9].

We used the method of activated magnetoresistivity for reconstruction of the 2D-hole gas spectrum under quantizing magnetic fields in p-Ge/Ge$_{1-x}$Si$_x$ systems with complex valence band structure [10]. Measurements of the longitudinal $r_{xx}$ and Hall $r_{xy}$ resistivities have been carried out in magnetic fields up to 12T at $T = (1.7 \div 15)$K (Fig. 1).

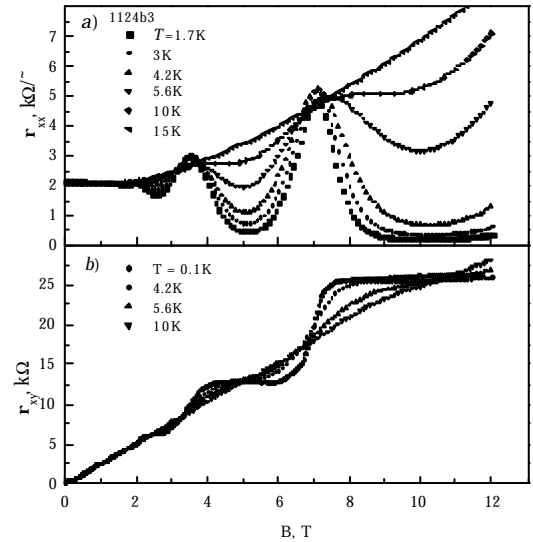

Figure 1. Longitudinal (a) and Hall (b) magnetoresistivity of the sample 1124b3 for the different temperatures.

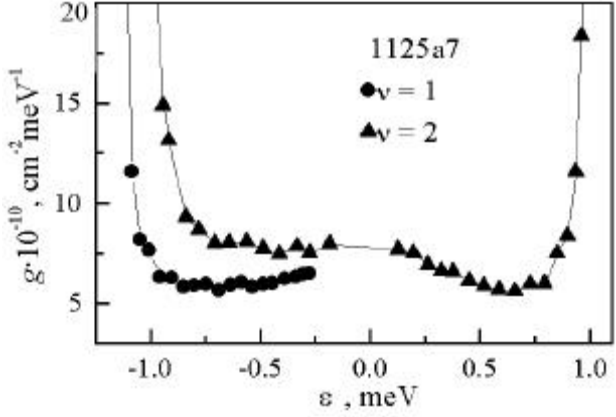

Figure 2. The background DOS for the sample 1125, as deduced from the activation energy. $\varepsilon = 0$ corresponds to the middle of an energy interval between two LL's.

The following results for the mobility gap DOS as a function of energy $g(\varepsilon)$ have been obtained. Even in the middle of a gap when the filling factor is close to an integer, the density of localized states is found to have values comparable with the DOS of 2DHG without magnetic field ($g_0 \cong 4.5 \cdot 10^{10}$ cm$^{-2}$meV$^{-1}$). Moreover, $g(\varepsilon)$ remains almost constant in the overwhelming part of the energy intervals between adjacent LL: $g(\varepsilon) \cong g_c = (5 \div 7) \cdot 10^{10}$ cm$^{-2}$meV$^{-1}$ for $\nu = 1$ and $\nu = 2$ (Fig. 2). The method of $g_c$ value estimation uses the experimentally obtained values of the conductivity activation energy as a magnetic field function $\varepsilon_A(B)$ (see Ref. [10] for the details). This is a rather rough method, but it allows us to find the values of $g_c$ correct in the order of magnitude, although somewhat overestimated. The main conclusion is that $g_c(\varepsilon)$ is practically constant and comparable with $g_0$ within the energy interval between adjacent LL's. This result is consistent qualitatively with the data for structures with $n$-type conductivity [4, 5, 8, 9]. As for our value of $g_c$, it is about an order of magnitude higher than those for InGaAs/InP [4] and for high-mobility AlGaAs/GaAs [5] heterostructures but comparable with those for Si-MOSFET [8] and intermediate-mobility AlGaAs/GaAs heterostructures [9].

As all the short-range impurity potential models lead to an exponential drop in DOS between Landau levels, the clear picture for the DOS in QHE regime may be presented only in terms of the long-range potential fluctuations in combination with the oscillating dependence of DOS on the filling factor. Such an idea has been advanced in the early work of Shklovskii and Efros [11] and later developed in series of works of Efros et al. (see [12, 13] and references therein). In selectively doped heterostructures, the smooth random potential is formed by fluctuations in concentration of remote impurities.

For a random potential $V(r)$, smooth on the scale of magnetic length $l_B$, the localization in QHE regime can be discussed in terms of semiclassical quantization and percolation [14]. In the quasiclassical limit, the electron energy in quantizing magnetic field may be presented as

$$E_N(r_0) = \hbar \omega_c (N + 1/2) + V(r_0) \qquad (1)$$

with $r_0$ being the oscillator center coordinate. Thus the smooth potential removes the degeneracy on $r_0$ and makes the LL energy dependent on spatial coordinates (Fig. 3).

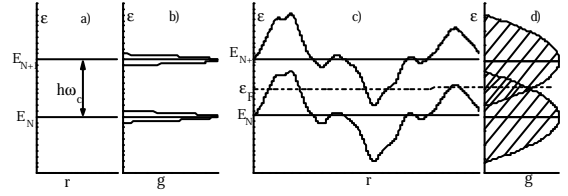

Figure 3. The DOS at the LL's in the absence of scattering (a) and with scattering (b). Spatial smooth random potential relief of the LL's (c) and an appropriate picture of the extended DOS at the LL's (d).

### 3.1.2. Impurity potential parameters for nearly integer filling factors

We report here an order of magnitude evaluation of the spatial scale and amplitude of random potential in p-Ge/Ge$_{1-x}$Si$_x$ heterostructures in QHE regime obtained from an analysis of the mobility gap DOS. Two models for random impurity potential were used.

i) The model with randomly distributed charged centers located within a thick layer close to the 2D-electron (hole) gas [11], for which the relation between fluctuation amplitude $F$ and scale $L$ reads:

$$F(L) = \beta \frac{e^2 \sqrt{NL}}{æ}, \qquad (2)$$

$\beta$ is a numerical coefficient ($\beta \cong 0.1$ [12]), $N$ – the density of charged impurities (per volume) and $æ$ – static dielectric constant.

ii) The model of the system with a spacer: a condenser with 2D electron (hole) gas as one plate and randomly distributed charged centers as the other plate, separated by a distance $d_s$ [12,13]. In this case:

$$F(L) = \frac{e^2 \sqrt{2\pi C}}{æ} \sqrt{\ln \frac{L}{2d_s}}, \qquad (3)$$

where $C$ is the average impurity density (per area).

It is seen from Eqs. (2) and (3) that without screening the amplitude $F$ diverges at large $L$. When the filling factor is close to an integer ($i$) very small concentration of electrons $\delta n \ll n_B$ can be redistributed in space and thus one occurs in conditions of so called nonlinear screening [11-13] ("threshold" screening in terms of [15]). For $\nu = i$ exactly, the screening is realized only due to electrons (and holes) induced by an overlap of adjacent fluctuating Landau levels, so the amplitude of random potential is of the order of corresponding LL gap.

For the investigated heterostructures $N \cong 10^{17}$ cm$^{-3}$ ($C = Nd_a \cong 10^{11}$ cm$^{-2}$) and the mean distance between impurities $N^{-1/3} \cong 200$Å is comparable both with the width of 2D Ge layer $d_w \cong 200$Å and the width of doped part of the sample $d_a \cong 100$Å. Thus the described models are not valid precisely but they are suitable to obtain a range of random potential parameter values.

In the nonlinear screening regime, we have the DOS in the middle of mobility gap [11-13] of width $W \cong 2$meV [10] for the two models, respectively:

i)
$$g(W/2) = \frac{4be^2 N}{æ W^2} \cong 7.5 \cdot 10^{10} \text{cm}^{-2}\text{meV}^{-1}, \quad (4)$$

ii)
$$g(W/2) = \frac{2\sqrt{C}}{7 W d_s} \cong 9.5 \cdot 10^{10} \text{cm}^{-2}\text{meV}^{-1}. \quad (5)$$

So, without any fitting parameter we obtain a rather reasonable evaluation of background DOS, and the two models yield values close to each other. For random potential amplitude comparable to the mobility gap, $F \cong W$, we obtain an evaluation of the nonlinear screening length $L_c$ (the scale of optimal fluctuation): $L_c \cong 1000$Å for model (i) (see Eq.(2)) and $L_c \cong 400$Å for model (ii) (see Eq.(3)). As seen in both cases the spatial scale of fluctuations is essentially larger than the magnetic length ($l_B \cong 80$Å at $B = 10$T), hence the random potential may be really regarded as the smooth one.

Thus, an order of magnitude evaluations of the random impurity potential parameters for the p-Ge/Ge$_{1-x}$Si$_x$ heterostructures indicate that in the vicinity of integer filling factors $n = 1$ and $n = 2$ (i.e. in the regions of the QHE plateaux) a sharp broadening of LL takes place (Fig. 3). It is reputed that for the filling factor close to a half-integer (the regions of plateau to plateau transition) the potential fluctuations would be small due to effective (linear) electron screening [11-13].

## 3.2. The QHE plateau-to-plateau transition regions

### 3.2.1. The width of the extended state band

The QHE regime may be regarded as a sequence of quantum phase insulator-metal-insulator transitions when the DOS of 2D system in quantizing magnetic fields is scanned by the Fermi energy. In terms of this conception the transition regions between adjacent QHE plateaux, as well as the width of appropriate $r_{xx}(B)$ peaks, should get more and more narrow as the temperature approaches zero. In the theoretical framework of scaling (see, for example, [16] and reference therein) the width of the transition regions goes to zero as

$$dB_{i \to (i+1)} \sim T^k \quad (6)$$

where $k = 1/zn$, $n = 7/3$ is the critical index of localization length and $z = 1$ is the dynamical critical index.

The pioneer experimental study of Wei et al. [17] on low mobility InGaAs/InP heterostructures has strongly supported the power law behavior of Eq.(6). The evolution of the width of the $r_{xx}$ peaks and of the inverse maximal slope of the $r_{xy}$ steps, $(dr_{xy}/dB)_{max}^{-1}$, as a function of temperature corresponds to (6) with nearly universal value of exponent $k = 0.4 \pm 0.04$ for several LLs. The scaling behavior with $k = (0.42 \div 0.46)$ has been reported later for QHE plateau-to-plateau transition in GaAs/AlGaAs heterostructures [18] and in p-SiGe quantum wells [19] and for QHE-to-insulator transition in GaAs/AlGaAs [18] and InGaAs/InP heterostructures [20].

In other series of experimental works the universality of exponent $k$ was questioned (see references in review article [14]). For instance, the measured values of κ increased from 0.28 to 0.81 with decreasing mobility in AlGaAs/GaAs heterostructures [21] or the values of $k$ between 0.2 and 0.65 were obtained for six subbands of Si-MOSFETs [22].

In a recent work of Shahar et al. [23] a novel transport regime distinct from the critical scaling behavior was reported to exist asymptotically close to the transition at very low temperatures. Studying the QHE-to-insulator transition in a variety of GaAs/AlGaAs and InGaAs/InP samples at temperatures down to 70mK, they found an exponential dependence of $r_{xx}$ on filling factor on the both sides of the critical FF value $n_c$ ($\Delta n = |n - n_c|$):

$$r_{xx} = \exp(-\Delta n / n_0(T)) \quad (7)$$

and emphasized that the effective transition width $n_0(T)$ appears to vary as $\alpha' T + \beta$ rather than to exhibit $T^k$ scaling behavior. It means that even at $T = 0$ the transition is of a finite width unless a different conduction mechanism takes over at still lower temperatures. The authors noted that some of their InGaAs/InP samples were from the same growth as the sample in Ref [17] and that they also revised their own previous data for GaAs/AlGaAs samples [18].

To estimate the width of the band of delocalized states in our Ge/Ge$_{1-x}$Si$_x$ samples we have analyzed magnetoresistance data in transition region between the first and second QHE plateaux in two ways. First, we used the description of $s_{xy}(B)$ dependences in terms of so-called scattering parameter [24]

$$s = \exp(-\Delta n / n_0(T)). \quad (8)$$

For $1 \to 2$ plateau-to-plateau transition the scattering parameter can be extracted according to [19, 25]

$$s_{xy} = 2 - s^2/(1+s^2). \quad (9)$$

The other way we used was to find the maximum slope of $(dr_{xy}/dB)_{max}$ in a transition region and to draw the inverse of it in reliable units against the temperature as that in Ref [17].

In Fig. 4 and 5 the $s_{xy}(B)$ and $s(n)$ dependences for one of the investigated samples (1124b$_3$) are presented. Figures 6a,b depict $n_0(T)$ dependences extracted according to Eq. (8) in a log-log graph and on a linear scale. It is seen from Fig. 6a that the data cannot be satisfactorily described by a power law $n_0 \sim T^k$ (it is not a straight line on a log-log plot). On the other hand the data are far more compatible with a linear dependence

$$n_0(T) = \alpha T + \beta \quad (10)$$

with $\alpha = 0.076$, $\beta = 0.027$ and $\beta/\alpha = 2.8$ K (Fig. 6b).

The frontal treatment of the data by the inverse maximum slope of $r_{xy}(B)$ yields qualitatively the same but slightly less accurate result of Eq.(10) with $\beta/\alpha = 2.6$ K for sample $1124b_3$ and $\beta/\alpha = 2.3$ K for sample $1123a_6$ (Fig. 7).

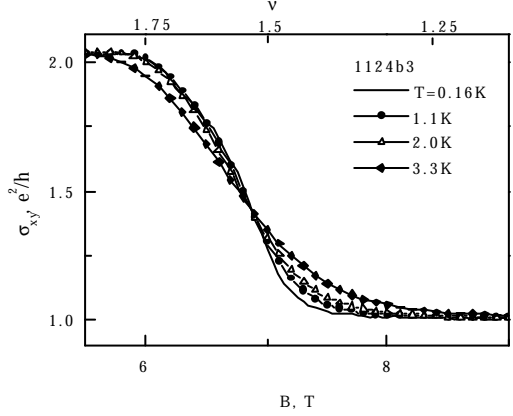

Figure 4. Hall conductivity, plotted as a function of filling factor $n$

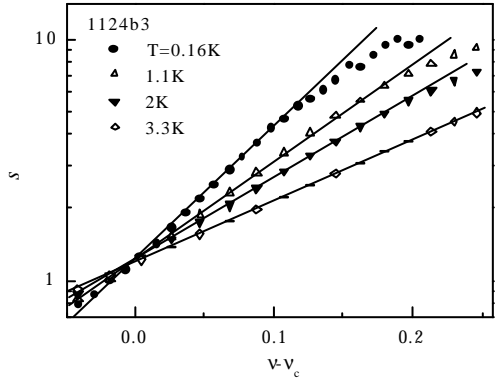

Figure 5. Scattering parameter $s$, defined in the text, derived from the $s_{xy}$ data shown in Fig. 4 for the sample 1124b3.

As is pointed out in [23] the ratio $\beta/\alpha$ defines a temperature $T^*$ that is founded to be characteristic of the material system. So, $T^*$ occurred to be close to 0.5K for InGaAs/InP samples and 50mK for GaAs/AlGaAs samples [23]. It is seen that for Ge/GeSi samples studied here the characteristic temperature is about 2.5K (2.3 ÷ 2.8 K).

In the theoretical work of Pruisken et al. [26] and in experimental work of van Schaijk et al. [20] it is emphasized an essential importance of short range random potential scattering for studying scaling phenomena as the long-range potential fluctuations dramatically complicate the observability of the critical phenomenon. In their opinion the linear behavior ($n_0 = \alpha T + \beta$) is semiclassical in nature and should be observed at finite $T$ and in samples with predominantly slowly varying potential fluctuations.

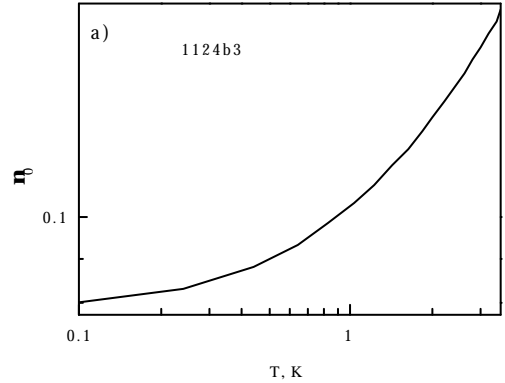

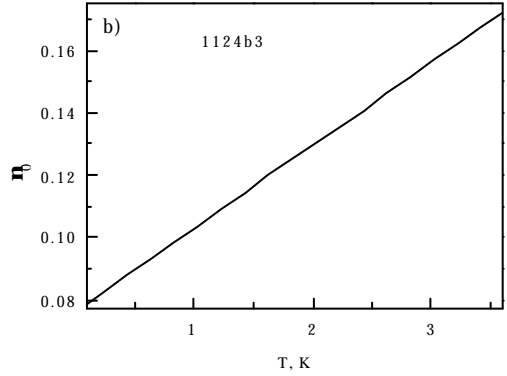

Figure 6. a) A log-log graph of $n_0(T)$ of Eq.8, plotted against $T$. b) Same as (a), plotted using a linear graph. Solid lines are the best fit.

The most simple and natural reason of the linear $n_0(T)$ dependence, namely, the thermal broadening of a quantum critical phase transition, is suggested and confirmed by calculation in the work of Coleridge and Zawadzki [25]. It is shown there that the thermal broadening not only yields the linear increase of $n_0(T)$ but also a temperature dependent *increase* of the $s_{xx}$ peak height as the temperature is lowered. And quite so it is in their experiment.

There is nothing about the temperature dependence of $r_{xx}$ (or $s_{xx}$) peak value in the work of Shahar et al. [23]. But we observe the linear $n_0(T)$ dependence in Ge/GeSi samples in that temperature interval where the peak values of $s_{xx}$ undoubtedly *decrease* with the lowering of $T$ (see section 3.2.2). Then we are not in the conditions of thermal broadening, in contrast to the experiment [25].

We think that the answer on the main question about the finite $T \to 0$ width of QHE transitions may be found in the works treating the influence of Coulomb interactions on the screening of smooth disorder potential [27, 28, 29]. The theory includes screening within Thomas-Fermi approximation appropriate for a smooth disorder.

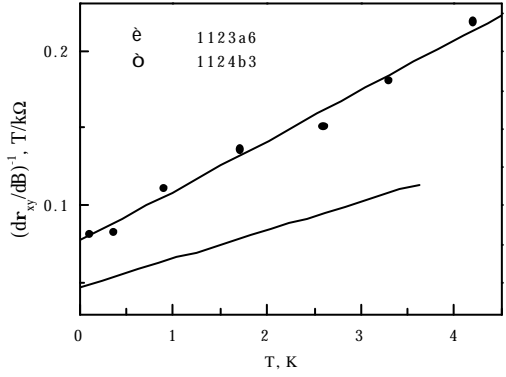

Figure 7. The inverse maximum slope of $(d\rho_{xy}/dB)^{-1}_{max}$ as a function of temperature for two samples 1123a6 and 1124b3.

The effect of electron-electron interaction manifests in that the regions of the third kind occur in the sample in addition to the local areas of full and empty LL present in the noninteracting system (Fig. 8). The new "metallic" regions are ones in which the local electron density is between zero and that of the full Landau level. Then the percolation description must be revised as the metallic region percolates through the sample over a finite range of magnetic field near the critical value. One therefore expects transition between Hall plateaus to have a finite width in filling factors even in the low-temperature limit.

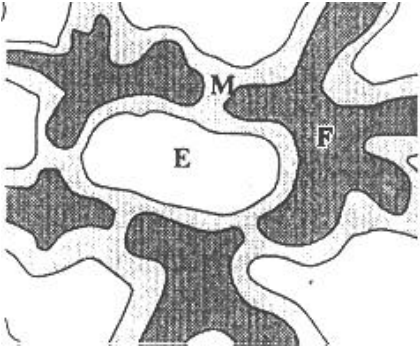

Figure 8. Percolation via metallic regions at $n < n_c$. Areas in which the LL is locally empty, partially occupied ("metallic"), and full are denoted be E, M, and F, respectively (after [29]).

### 3.2.2. Conductivity on delocalized states

In Fig. 9 the typical $\rho_{xx}(B)$ and $\rho_{xy}(B)$ dependences for our heterostructures at $T \cong 0.1K$ are presented. In the QHE regime we have observed that at $T \geq (3\div 4)K$ the amplitude of $\rho_{xx}$ peak diminishes as the temperature is lowered (Fig. 10). The decreasing of peak value $\rho_{xx}^{peak}$ at $T \to 0$ is caused by decreasing of peak value $\sigma_{xx}^{peak}$ as in high magnetic field $\rho_{xx} = \sigma_{xx}/\sigma_{xy}^2$ and peak of $\rho_{xx}$ corresponds to $\sigma_{xy} = (i+\frac{1}{2})e^2/h$, irrespective of temperature. Such a behavior of 2D-system conductivity in QHE regime, observed first by Wei et al. [30, 31] on InGaAs/InP system, was treated by them as a manifestation of a scaling regime.

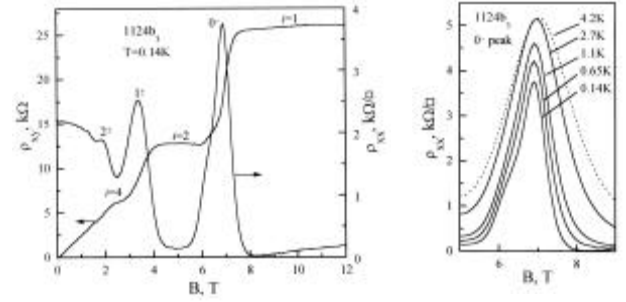

Figure 9. Longitudinal and Hall magnetoresistivity for sample 1124b3 at $T = 0.14$ K.

Figure 10. The $0^-$ $\rho_{xx}^{peak}$ value for the same sample (Fig. 9) at different temperatures.

The expression for conductivity at finite temperatures can be written as [32]

$$\sigma_{xx}(T) = -\int \sigma_{xx}(E) \frac{\partial f(E-E_F)}{\partial E} dE \quad (11)$$

where $f(E - E_F)$ is the Fermi-Dirac distribution function and $\sigma_{xx}(E)$ is the partial contribution to the dissipative conductivity of states with energy $E$. If $E = E_c$ is the critical energy in the LL-center and $\Gamma$ is the width of the band of delocalized states, only the states in the energy interval $|E - E_c| \leq \Gamma$ contribute to the conduction process. Then we present the partial conductivity as [33]

$$\sigma_{xx}(E) = \sigma_c \frac{\Gamma^2}{(E-E_c)^2 + \Gamma^2}. \quad (12)$$

For $E_F = E_c$ one has from Eq. (11) and (12):

$$\sigma_{xx}^{peak}(T) = \frac{\pi}{4}\frac{\Gamma}{kT}\sigma_c, \quad (\Gamma < kT)$$
$$\sigma_{xx}^{peak}(T) = \sigma_c \qquad (\Gamma > kT) \quad (13)$$

Hence, two regions in temperature may be distinguished for the behavior of the $\sigma_{xx}$ peak amplitudes. In the low-temperature region $kT \ll \Gamma$ the value of $\sigma_{xx}^{peak}$ is completely determined by the conduction mechanism inherent to the band of delocalized states. When $kT > \Gamma$ the thermal smearing of the Fermi step becomes the main factor in accordance with the analysis of Coleridge and Zawadzki [25]. The fraction of extended states equal to $\Gamma/kT$ decreases with the temperature increase, that leads to the decreasing of $\sigma_{xx}^{peak}(T)$. The maximum $\sigma_{xx}^{peak}(T)$ should be reached at $kT \cong \Gamma$.

Let us examine the experimental results. Fig. 11 illustrates a nonmonotonic temperature dependence of $\sigma_{xx}^{peak}$ for the two of investigated samples. It appears reasonable to think that the temperature region in which the amplitude of the peak begins to decrease with $T$-lowering corresponds to the conductivity just on the band of extended states, i.e. to condition $kT < \gamma$. For all the samples studied here the $\sigma_{xx}^{peak}(T)$ dependence is close to linear at sufficiently low temperatures $T < (2 \div 2.5)$ K (Fig. 12a,b). It is remarkable that these $T$-values are correlated with the characteristic values of $T^*$ defined as the ratio $b/a$ in the linear $T$-dependence of the QHE plateau-plateau transition width (see section 3.2.1). Thus we have a reason to suppose that the conditions $kT < \gamma$ and $T < T^*$ are closely linked to each other.

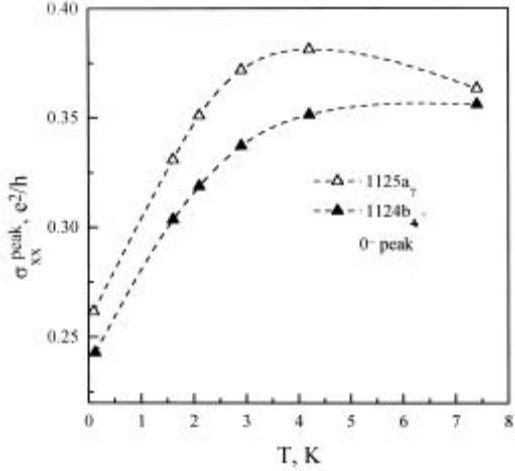

Figure 11. The temperature dependence of $\sigma_{xx}^{peak}(T)$ for two samples.

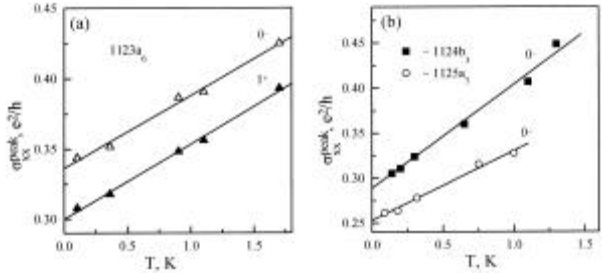

Figure 12. The temperature dependence of $\sigma_{xx}^{peak}(T)$ for different peaks and samples: (a) 1123a6 $0^-$ and $1^+$ peaks and (b) $0^-$ peak for sample 1124b3 and 1125a7.

The linear extrapolation of $\sigma_{xx}^{peak}(T)$ dependences to $T = 0$ yields the limiting values of conductivity on extended states in the center of $0^-$ or $1^+$ Landau subbands $\sigma^* = (0.26 \div 0.33)e^2/h$ (see Fig. 12). The $\sigma^*$ values obtained are extremely close to values of $\sigma_{xx}$ for $0^-$ and $1^+$ peaks at $T = 50$ mK for the InGaAs/InP heterostructures $\sigma_{xx}^{peak} = (0.25 \div 0.35)e^2/h$ [31].

A two-parameter scaling theory of Pruisken for the case of *short-range disorder* [34, 35] yields the value $\sigma^* < e^2/\pi h$ for limiting value of $\sigma_{xx}^{peak}$ at $T \to 0$ as well as allows a power law for the temperature dependence of $\Delta \sigma = (\sigma_{xx}^{peak} - \sigma^*)$ [33, 36]. But all the theories of the quantum phase transitions in the QHE regime in a *smooth disorder potential* [24, 26] as well as numerical simulations in the framework of network model [14, 37] for *noninteracting electrons* severely predict a universal value of $e^2/2h$ for peak value of the conductivity between QHE plateaux. This prediction is rather badly confirmed by experiment as is pointed out in recent review articles [38, 39]. For inter-QHE-plateaux transitions, most researchers report critical amplitude $\sigma_{xx}^{peak}$ that is not only significantly $(40 \div 80\%)$ smaller than the theoretically expected value but in many cases is also $T$-dependent. The violation of the universality of $\sigma_{xx}^{peak}$ value as well as of the scaling behavior of the peak width may be caused by the electron-electron interaction in the screening of smooth potential [27-29].

## 4. Concluding remarks

In selectively doped p-Ge/Ge$_{1-x}$Si$_x$ heterostructures investigated here the main scattering mechanism for quasi-2D holes in Ge quantum wells at low temperatures is the scattering on remote ionized boron impurities located in the Ge$_{1-x}$Si$_x$ barriers. The fluctuations in the density of randomly distributed remote impurities act as a source of the smooth disorder potential causing localization effects in the quantum Hall regime. Screening of this disorder takes on a very different character depending on the value of the filling factor. When the Fermi level is near the center of Landau subband (half-integer FFs) electrons are free to adjust their density and screening of random potential is good, but when it lies in the mobility gaps between LL's (nearly integer FF's) they cannot and screening is poor.

Only in the framework of disorder potential, smooth on the scale of the magnetic length, it occurs possible for us to explain unusually high values of background DOS obtained from the analysis of thermally activated magnetoresistance in the QHE plateaux regions in the vicinity of $\nu = 1$ and $\nu = 2$. In the models with nonlinear screening of long-range random impurity potential we obtain a reasonable estimation both for the density of localized states and for spatial scale of potential fluctuation, which really occurs to be rather large as compared to magnetic length.

On the other hand, for half-integer FF's, the linear temperature dependence of the effective QHE plateau-to-plateau transition width $\nu_0(T) = \beta + \alpha T$ is observed in our Ge/Ge$_{1-x}$Si$_x$ samples in contrast to scaling behavior inherent to systems with shot-range disorder. This result is in accordance with the data of recent experimental work [23] for other semiconductor systems. It is tempted to consider the finite width of the QHE transition, even at $T \to 0$, as a consequence of an effective screening of smooth random potential owing to Coulomb repulsion of electrons [27-29].

The "metallic" regions percolating over a range of FF's in the vicinity of half-integer $n = n_c$ should be formed in the plane of the sample due to influence of electron-electron interaction. The mechanism of conductivity on the "metallic" band is not clear [29] but we argue that the reason of the observed (linear) $T$-dependence of the amplitude of $r_{xx}$ ($\sigma_{xx}$) peak at $T < T^* = \beta/\alpha$ may be the same as that for $n_0(T)$.

**Acknowledgments**

This work was supported by Russian Foundation for the Basic Researches, projects 99-02-16256-a, 01-02-17685-a, 01-02-06131-mas, 01-02-0625-mas and 6-th concurs-expertise RAS (1999) No. 68.